\documentclass[11pt,a4paper]{article}

\usepackage{epsfig}
\usepackage{amstex}
\usepackage{amssymb}

\begin{document}

\newcommand{\zp}[3]{{\sl Z. Phys.} {\bf #1} (19#2) #3}
\newcommand{\np}[3]{{\sl Nucl. Phys.} {\bf #1} (19#2)~#3}
\newcommand{\pl}[3]{{\sl Phys. Lett.} {\bf #1} (19#2) #3}
\newcommand{\pr}[3]{{\sl Phys. Rev.} {\bf #1} (19#2) #3}
\newcommand{\prl}[3]{{\sl Phys. Rev. Lett.} {\bf #1} (19#2) #3}
\newcommand{\prep}[3]{{\sl Phys. Rep.} {\bf #1} (19#2) #3}
\newcommand{\fp}[3]{{\sl Fortschr. Phys.} {\bf #1} (19#2) #3}
\newcommand{\nc}[3]{{\sl Nuovo Cimento} {\bf #1} (19#2) #3}
\newcommand{\ijmp}[3]{{\sl Int. J. Mod. Phys.} {\bf #1} (19#2) #3}
\newcommand{\ptp}[3]{{\sl Prog. Theo. Phys.} {\bf #1} (19#2) #3}
\newcommand{\sjnp}[3]{{\sl Sov. J. Nucl. Phys.} {\bf #1} (19#2) #3}
\newcommand{\cpc}[3]{{\sl Comp. Phys. Commun.} {\bf #1} (19#2) #3}
\newcommand{\mpl}[3]{{\sl Mod. Phys. Lett.} {\bf #1} (19#2) #3}
\newcommand{\cmp}[3]{{\sl Commun. Math. Phys.} {\bf #1} (19#2) #3}
\newcommand{\jmp}[3]{{\sl J. Math. Phys.} {\bf #1} (19#2) #3}
\newcommand{\nim}[3]{{\sl Nucl. Instr. Meth.} {\bf #1} (19#2) #3}
\newcommand{\el}[3]{{\sl Europhysics Letters} {\bf #1} (19#2) #3}
\newcommand{\ap}[3]{{\sl Ann. of Phys.} {\bf #1} (19#2) #3}
\newcommand{\jetp}[3]{{\sl JETP} {\bf #1} (19#2) #3}
\newcommand{\jetpl}[3]{{\sl JETP Lett.} {\bf #1} (19#2) #3}
\newcommand{\acpp}[3]{{\sl Acta Physica Polonica} {\bf #1} (19#2) #3}
\newcommand{\vj}[4]{{\sl #1~}{\bf #2} (19#3) #4}
\newcommand{\ej}[3]{{\bf #1} (19#2) #3}
\newcommand{\vjs}[2]{{\sl #1~}{\bf #2}}
\newcommand{\hep}[1]{{\sl hep--ph/}{#1}}
\newcommand{\hex}[1]{{\sl hep--ex/}{#1}}
\newcommand{\desy}[1]{{\sl DESY-Report~}{#1}}
\catcode`@=11 \def\citer{\@ifnextchar
  [{\@tempswatrue\@citexr}{\@tempswafalse\@citexr[]}}

\def\@citexr[#1]#2{\if@filesw\immediate\write\@auxout{\string\citation{#2}}\fi
  \def\@citea{}\@cite{\@for\@citeb:=#2\do
    {\@citea\def\@citea{--\penalty\@m}\@ifundefined {b@\@citeb}{{\bf
          ?}\@warning
       {Citation `\@citeb' on page \thepage \space undefined}}%
\hbox{\csname b@\@citeb\endcsname}}}{#1}}
\catcode`@=12
\marginparwidth 3cm
\setlength{\hoffset}{-1cm}
\newcommand{\mpar}[1]{{\marginpar{\hbadness10000%
                      \sloppy\hfuzz10pt\boldmath\bf\footnotesize#1}}%
                  \typeout{marginpar: #1}\ignorespaces}
\def\mda{\mpar{\hfil$\downarrow$\hfil}\ignorespaces}
\def\mua{\mpar{\hfil$\uparrow$\hfil}\ignorespaces}
\def\mla{\marginpar[\boldmath\hfil$\rightarrow$\hfil]%
                   {\boldmath\hfil$\leftarrow $\hfil}%
                   \typeout{marginpar:
                     $\leftrightarrow$}\ignorespaces}
\renewcommand{\abstractname}{Abstract}
\renewcommand{\figurename}{Figure}
\renewcommand{\thefootnote}{\fnsymbol{footnote}}

\providecommand{\SP}{\scriptstyle}
\providecommand{\ns}{\hspace{2mm}}

\providecommand{\stl}{\tilde{t}_L}
\providecommand{\str}{\tilde{t}_R}
\providecommand{\ste}{\tilde{t}_1}
\providecommand{\stz}{\tilde{t}_2}
\providecommand{\stez}{\tilde{t}_{1,2}}
\providecommand{\sti}{\tilde{t}_i}
\providecommand{\stj}{\tilde{t}_j}
\providecommand{\stk}{\tilde{t}_k}
\providecommand{\st}{\tilde{t}}
\providecommand{\sql}{\tilde{q}_L}
\providecommand{\sqr}{\tilde{q}_R}
\providecommand{\sq}{\tilde{q}}
\providecommand{\gt}{\tilde{g}}

\providecommand{\stlb}{\bar{\tilde{t}}_L}
\providecommand{\strb}{\bar{\tilde{t}}_R}
\providecommand{\strlb}{\bar{\tilde{t}}_{R,L}}
\providecommand{\steb}{\bar{\tilde{t}}_1}
\providecommand{\stzb}{\bar{\tilde{t}}_2}
\providecommand{\stezb}{\bar{\tilde{t}}_{1,2}}
\providecommand{\stib}{\bar{\tilde{t}}_i}
\providecommand{\stjb}{\bar{\tilde{t}}_j}
\providecommand{\stkb}{\bar{\tilde{t}}_k}
\providecommand{\stb}{\bar{\tilde{t}}}
\providecommand{\sqb}{\bar{\tilde{q}}}
\providecommand{\tb}{\bar{t}}

\providecommand{\stn}{\tilde{t}_0}
\providecommand{\stln}{\tilde{t}_{L 0}}
\providecommand{\strn}{\tilde{t}_{R 0}}
\providecommand{\sten}{\tilde{t}_{1 0}}
\providecommand{\stzn}{\tilde{t}_{2 0}}
\providecommand{\stin}{\tilde{t}_{i 0}}
\providecommand{\stjn}{\tilde{t}_{j 0}}
\providecommand{\chijn}{\tilde{\chi}_j^0}
\providecommand{\chijp}{\tilde{\chi}_j^+}

\providecommand{\mse}{m_{\tilde{t}_{\SP 1}}}
\providecommand{\msz}{m_{\tilde{t}_{\SP 2}}}
\providecommand{\msi}{m_{\tilde{t}_{\SP i}}}
\providecommand{\msj}{m_{\tilde{t}_{\SP j}}}
\providecommand{\msk}{m_{\tilde{t}_{\SP k}}}
\providecommand{\msen}{m_{\tilde{t}_{\SP 1 0}}}
\providecommand{\mszn}{m_{\tilde{t}_{\SP 2 0}}}
\providecommand{\msjn}{m_{\tilde{t}_{\SP j 0}}}
\providecommand{\mst}{m_{\tilde{t}}}
\providecommand{\ms}{m_{\tilde{q}}}
\providecommand{\mg}{m_{\tilde{g}}}
\providecommand{\mt}{m_t}

\providecommand{\tmixn}{\tilde{\theta}_0}
\providecommand{\tmix}{\tilde{\theta}}
\providecommand{\delZ}{\delta \tmix}

\providecommand{\ZMM}{Z^{-1/2}}
\providecommand{\ZM}{Z^{1/2}}
\providecommand{\MM}{{\cal M}^2}

\providecommand{\as}{\alpha_s}
\providecommand{\msbar}{\overline{{\rm MS}}}
\providecommand{\drbar}{\overline{{\rm DR}}}
\providecommand{\mssm}{{\cal MSSM}}

\providecommand{\real}{{\cal\text{Re\,}}}
\providecommand{\imag}{{\cal\text{Im\,}}}
\providecommand{\sig}{\rm{sig\,}}

\providecommand{\ra}{\rightarrow}
\providecommand{\lra}{\leftrightarrow}
\providecommand{\Ra}{\Rightarrow}
\providecommand{\PL}{\frac{1}{2}(1-\gamma_5)}
\providecommand{\PR}{\frac{1}{2}(1+\gamma_5)}
\providecommand{\PLR}{{\rm P_{L,R}}}
\providecommand{\Pe}{{\rm P_1}}
\providecommand{\Pz}{{\rm P_2}}
\providecommand{\Pez}{{\rm P_{1,2}}}
\providecommand{\PX}{{\cal P}_{12}}
\providecommand{\ce}{\cos \tmix}
\providecommand{\se}{\sin \tmix}
\providecommand{\cz}{\cos(2\tmix)}
\providecommand{\sz}{\sin(2\tmix)}
\providecommand{\czsq}{\cos^2(2\tmix)}
\providecommand{\szsq}{\sin^2(2\tmix)}

\renewcommand{\thefootnote}{\fnsymbol{footnote}}
\begin{titlepage}

\begin{flushright}
DESY 97-43 \\
CERN-TH/97-177\\
RAL-TR-97-056 \\
\end{flushright}

\vspace{1cm}

\begin{center}
\baselineskip25pt

\def\thefootnote{\fnsymbol{footnote}}
{\Large\sc Stop production at hadron colliders}

\end{center}

\setcounter{footnote}{3}

\vspace{1cm}

\begin{center}
\baselineskip12pt

{\sc 
W.~Beenakker$^1$\footnote{Research supported by a fellowship of the Royal 
                          Dutch Academy of Arts and Sciences.},
M.~Kr\"amer$^2$,  
T.~Plehn$^3$,
M.~Spira$^4$,
and P.M.~Zerwas$^3$} 

\vspace{1cm}

$^1$ Instituut--Lorentz, University of Leiden, The Netherlands 
\\[5mm]
$^2$ Rutherford Appleton Laboratory, Chilton, Didcot, OX11 0QX, UK 
\\[5mm]
$^3$ Deutsches Elektronen--Synchrotron DESY, D--22603 Hamburg, FRG
\\[5mm]
$^4$ CERN, Theory Division, CH--1211 Geneva 23, Switzerland

\vspace{0.3cm}

\end{center}

\vspace{2cm}

\begin{abstract}
  \normalsize \noindent Stop particles are expected to be the lightest
  squarks in supersymmetric theories and the search for these
  particles is an important experimental task. We therefore present
  the cross sections for the production processes $p\bar{p}/pp \to
  \ste\steb$ and $\stz\stzb$ at Tevatron and LHC energies in
  next-to-leading order supersymmetric QCD. The corrections stabilize
  the theoretical predictions for the cross sections, and they are
  positive, thus raising the cross sections to values above the
  leading-order predictions. Mixed $\ste\stzb/\steb\stz$ pairs can
  only be generated in higher orders at strongly suppressed rates.
\end{abstract}

\end{titlepage}

\def\thefootnote{\arabic{footnote}} \setcounter{footnote}{0}

\setcounter{page}{2}

\section[]{Introduction} 

Within the squark sector of supersymmetric theories, the top-squark
(stop) eigenstate $\ste$ is expected to be the lightest particle
\cite{ellis}. If the scalar masses in grand unified theories, for
instance, are evolved from universal values at the GUT scale down to
low scales the $\ste$ top-squark drops to the lowest value in the
squark spectrum. Moreover, the strong Yukawa coupling between top/stop
and Higgs fields gives rise to large mixing, leading to a potentially
small mass eigenvalue for $\ste$ \cite{ohmann}.

In $e^+e^-$ and $p\bar{p}/pp$ collisions stop particles are produced
in pairs. Present limits from LEP2 indicate a $\ste$ mass in excess of
67~GeV, independent of the mixing angle in the stop sector, but
depending on the lightest neutralino mass \cite{lep}.  Preliminary
analyses at the Tevatron have led to a lower limit of 93 GeV for the
$\ste$ mass, depending on the lightest neutralino mass
\cite{tevatron}.  \bigskip

The cross sections for the production of squarks ($\sq$) and gluinos
($\gt$) in hadron collisions have been calculated at the Born level
already quite some time ago \cite{born}. Only recently have these
theoretical predictions been improved by calculations of the
next-to-leading order (NLO) SUSY-QCD corrections for squark/gluino
production, with the final-state squarks restricted to the
light-flavor sector ($\sq\ne\st$) \cite{roland}. In the present paper
we supplement this analysis by the corresponding analysis for the stop
sector
\begin{equation}
p\bar{p}/pp \to 
\ste \steb + X 
\qquad \text{and} \qquad 
\stz \stzb + X 
\label{eq_intro}
\end{equation}
in the $p\bar{p}$ collisions of the Tevatron and the $pp$ collisions
of the LHC. This NLO calculation is motivated by two requirements:
First, to stabilize the theoretical predictions for the cross sections
with respect to the renormalization and factorization scales, which
introduce spurious parameters into fixed-order calculations. And
second, to improve the accuracy of the theoretical predictions for the
cross sections.  Extending the light-flavor analysis to stop
production is a necessary step since the mixing effects must be taken
into account properly. At NLO [{\it i.e.}~${\cal O}(\alpha_s^3)$] the
production cross sections are still diagonal in the stop sector.  The
production of mixed $\ste\stzb/\stz\steb$ pairs is suppressed as the
cross section is of order $\alpha_s^4$.  Since the calculation is
rather involved but the rate is small, we have studied this process in
the limit of large gluino mass to exemplify the expected size of the
non-diagonal cross section.

The cross sections for the diagonal production of stop particles,
Eq.\,(\ref{eq_intro}), depend essentially only on the mass of the
produced  stop particles, $\mse$ or $\,\msz$. The dependence of
the cross sections on the other SUSY parameters, {\it i.e.}~the gluino
mass $\mg$, the masses of the  other squarks and the mixing angle
$\tmix$, is very weak since these parameters affect only the
higher-order corrections and are not relevant at leading order.  

The results for the diagonal production of stop particles, $\ste
\steb$ and $\stz\stzb$, will be discussed in the next section. The
size of the cross section for mixed-pair production $\ste \stzb$ and
 $\stz \steb$ will be estimated subsequently.  \bigskip

\section[]{Diagonal stop-pair production}
 
\begin{figure}[ht]
\begin{center}
\epsfig{file=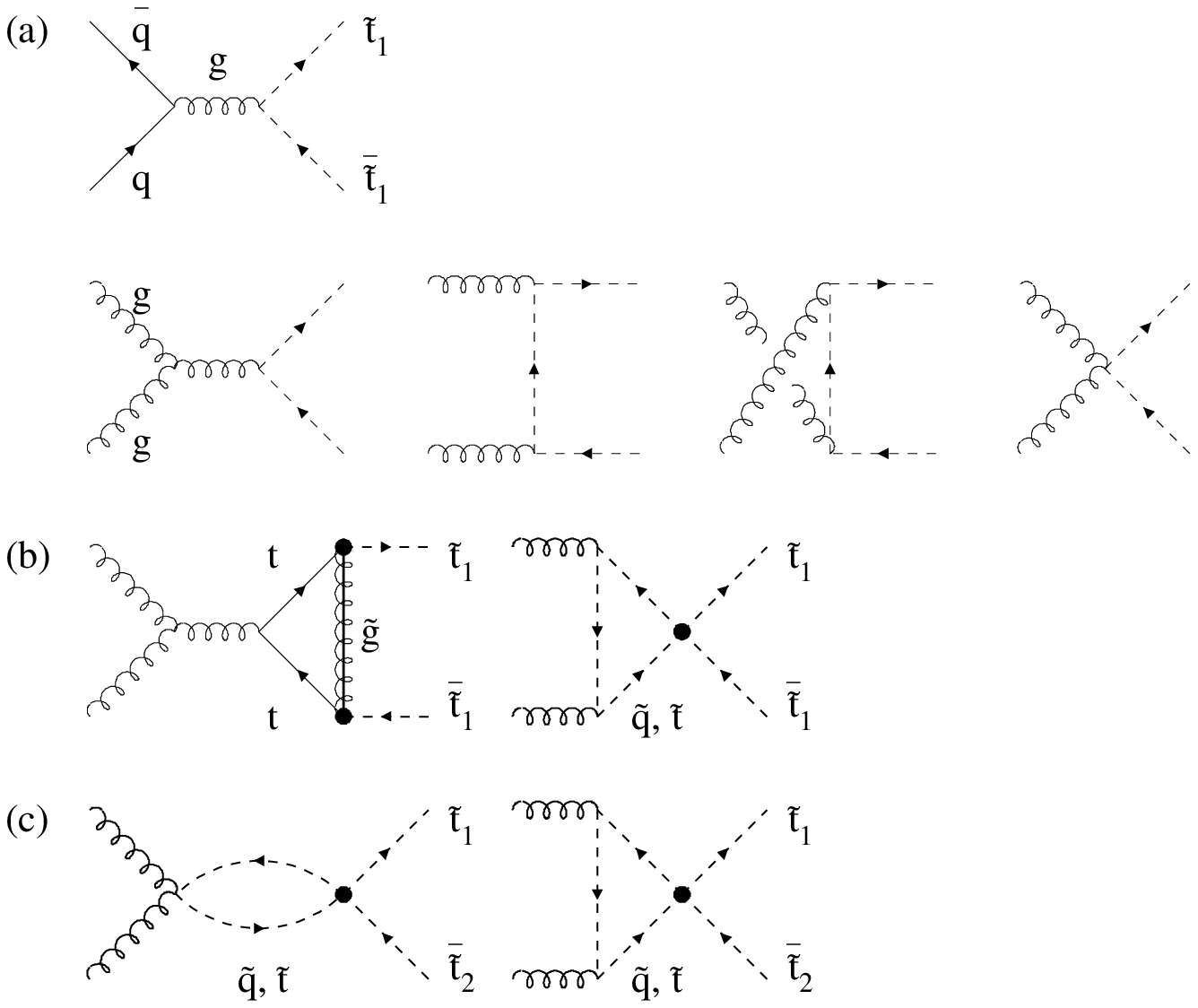,width=16cm}
\end{center}
\caption[]{\it Generic Feynman diagrams for the production of 
           pairs of stop
           particles: (a) Born diagrams for quark--antiquark 
           annihilation
           and gluon fusion; (b) higher-order 
           diagrams for the diagonal production 
           including stop mixing (dotted vertices); (c) non-diagonal 
           production in the limit of decoupled gluinos 
           (mixing vertices are dotted).\label{fig_feyn}}
\end{figure}

At hadron colliders, diagonal pairs of stop particles can be produced
at lowest order QCD in quark--antiquark annihilation and gluon--gluon
fusion:
\begin{eqnarray}
q \bar{q} &\rightarrow& \ste \steb 
\quad \text{and} \quad \stz \stzb \notag \\
g g       &\rightarrow& \ste \steb  
\quad \text{and} \quad \stz \stzb 
\label{eq_diagonal}
\end{eqnarray}
Mixed pairs $\ste \stzb$ and $\stz \steb$ cannot be produced in lowest
order  since the $g\st\st$ and $gg\st\st$ vertices are diagonal in
the chiral as well as in the mass basis. The relevant diagrams for the
reactions (\ref{eq_diagonal}) are shown in Fig.\,\ref{fig_feyn}a. The
corresponding cross sections for these partonic subprocesses may be
written as ($k=1,2$): 
\begin{alignat}{7}
&\hat{\sigma}_{LO}[q\bar{q}\to\stk\stkb] &\ =\ &  
    \frac{\alpha_s^2 \pi}{s} && \,\frac{2}{27}\,\beta_k^3 \\
&\hat{\sigma}_{LO}[gg\to\stk\stkb]       &\ =\ &
    \frac{\alpha_s^2 \pi}{s} && \,\left\{ 
        \beta_k \left( \frac{5}{48} + \frac{31\msk^2}{24s} \right)
      + \left( \frac{2\msk^2}{3s} + \frac{\msk^4}{6s^2} \right)
        \log\left(\frac{1-\beta_k}{1+\beta_k}\right) 
                                       \right\}
\label{eq_siglo}
\end{alignat}
The invariant energy of the subprocess is denoted by $\sqrt{s}$, the
velocity by $\beta_k=\sqrt{1-4 \msk^2/s}$. The cross sections coincide
with the corresponding expressions for light-flavor squarks in the
limit of large gluino masses, cf. Ref.\,\cite{roland}.\smallskip

The hadronic $p\bar{p}/pp$ cross sections are obtained by folding the
partonic cross sections with the $q\bar{q}$ and $gg$ luminosities.
At the Tevatron the dominant mechanism for large stop masses is
the valence $q\bar{q}$ annihilation. The fraction of $q\bar{q}$ events
rises from 0.55 to 0.86, if the $\ste$ mass is increased from 100 to
200~GeV. At the LHC the gluon-fusion mechanism plays a more prominent
role. For a $\ste$ mass below 200~GeV, more than 90\% of the events
are generated by $gg$ fusion, cf.~Table~\ref{tab_kfac}.

We closely follow the approach of Ref.\,\cite{roland} for the
calculation  of the NLO SUSY-QCD corrections.  The virtual ${\cal
  O}\left(\alpha_s\right)$ corrections involve the  usual SUSY-QCD
corrections to the propagators and vertices, as well as box and
rescattering diagrams, cf.~Ref.\,\cite{roland}. To this order
the mixing angle $\tmix$ enters the cross section only through
corrections involving the $t\st\gt$ and the four-squark couplings.
 The relevant couplings of this type are described by the
Lagrangeans
\begin{alignat}{7}
  &{\cal L}_3 \quad &=\ & - \sqrt{2}\,g_s T^a_{ij} 
   \left( 1 + \PX \right) 
      \bar{\gt}_a \Big[ \ce\,\PL -\se\,\PR \Big] t_j\,\ste{}_i^* 
      + \text{h.c.} \notag \\[1mm]
  &{\cal L}_4 \quad &=\ & - \frac{g_s^2}{8}\,\left( 1 + \PX \right)
      \ste{}_i^*\,\ste{}_j\,
      \bigg\{ \czsq\,A_2^{ijkl}\,\ste{}_k^*\,\ste{}_l 
            + 2\,\Big[ \szsq\,A_2^{ijkl}-A_1^{ijkl} \Big]\,\stz{}_k^*\,\stz{}_l
       \notag \\
  &&& \hphantom{- \frac{g_s^2}{8}\,\left( 1+\PX \right)\ste{}_i^*\,\ste{}_j aA}
          {}+ 4\,\cz\,A_1^{ijkl} \sum_{\sq\neq\st} 
              \big( \sql{}_k^*\,\sql{}_l - \sqr{}_k^*\,\sqr{}_l \big)
      \bigg\}   
  \label{L3L4} 
\end{alignat}
with the color tensors ($N_c$=3)
\begin{alignat}{5}
A_1^{ijkl} &= \delta^{il}\delta^{kj} - \delta^{ij}\delta^{kl}/N_c    
              \notag \\
A_2^{ijkl} &= \left( \delta^{il}\delta^{kj} + \delta^{ij}\delta^{kl} 
              \right) (N_c-1)/N_c
\end{alignat}
Here $\sq_{L/R}$ represent the left(L)- and right(R)-chirality
light-flavor squarks. Indices of the fundamental/adjoint
representation of color $SU(3)$ are denoted by $i,j,k$ and $a$,
respectively, and the generators of the fundamental representation by
$T^a$. The operator $\PX$ permutes the 1- and 2-components of the
top-squarks:
\begin{equation}
\PX \quad : \quad  [\ste \leftrightarrow \stz;\ \ce\to-\se,\ \se\to\ce]
\label{eq_perm}
\end{equation}
 Two typical diagams in which $\tmix$ enters the scattering
amplitude $gg \to \stk\stkb$ through vertex and rescattering
corrections are shown in Fig.\,\ref{fig_feyn}b. Since mixing enters
explicitly only through higher-order diagrams, the angle $\tmix$ need
not be renormalized \cite{angle} in the present calculation and it can
be identified with the lowest-order expression derived from the stop
mass matrix.

As usual, the virtual corrections are supplemented by gluon radiation
from color lines and vertices, as well as contributions from the
inelastic Compton processes $qg \to \stk\stkb q$ and $g\bar{q} \to
\stk\stkb \bar{q}$, cf.~Ref.\cite{roland} for diagrammatic details.
 \bigskip
 
 The singularities associated with the NLO corrections are isolated by
 means of dimensional regularization and renormalized within the
 $\msbar$ scheme.\footnote{Note that the spurious breaking of
   supersymmetry in the $\msbar$ scheme \cite{mv} has no effect on the
   NLO stop-pair cross sections. The Yukawa couplings become effective
   only as a part of the higher-order corrections and need not be
   renormalized.} The renormalization of the QCD coupling is performed
 in such a way that the heavy particles (top quarks, gluinos, and
 light-flavor squarks) are decoupled smoothly for momenta smaller than
 their masses. This implies that the heavy particles do not contribute
 to the evolution of the couplings and parton densities.  The masses
 of the light quarks ($q\neq t$) are neglected and the top-quark mass
 is set to $m_t=175$~GeV. For the calculation of gluon bremsstrahlung,
 the phase space for gluon radiation is split into two distinct
 regimes, one accounting for soft gluons and the other for the hard
 gluons. The separation is implemented by introducing a cut-off
 parameter $\Delta$ in the invariant mass of the radiated gluon and
 one of the heavy particles in the final state.  The cut-off parameter
 is chosen so small that it can be neglected with respect to any other
 mass scale in the process. As a result of the split-up of the phase
 space, terms of the form $\log^i \Delta$ $(i=1,2)$ occur in both the
 soft and the hard cross sections.  If soft and hard contributions are
 added up, any $\Delta$ dependence disappears from the cross sections
 in the limit $\Delta \to 0$.

At lowest order, the cross sections for diagonal $\ste\steb$ and
$\stz\stzb$ pair production are given by the same analytical
expression. At next-to-leading order,  the $t\st\gt$ and
four-squark interactions will affect the production cross sections for
the two diagonal pairs in different ways, introducing the explicit
dependence on the mixing angle. However, the $\tmix$ dependence will
turn out to be very mild. It follows from Eqs.\,(\ref{L3L4}), 
that the analysis of $\stz\stzb$ pairs can be copied from the analysis
of $\ste\steb$ pairs after applying the permutation $\PX$ defined in
Eq.\,(\ref{eq_perm}).

For the detailed description of the partonic cross-sections we
introduce scaling functions $f$,
\begin{equation}
\hat{\sigma}_{ij} = 
\frac{\as^2(\mu^2)}{\msk^2} 
\left\{ f_{ij}^{B}(\eta)
 + 4 \pi \as(\mu^2) \left[f_{ij}^{V+S}(\eta,r,\tmix) + f_{ij}^{H}(\eta) 
 + \bar{f}_{ij}(\eta) \log\left(\frac{\mu^2}{\msk^2}\right) \right] 
\right\}
\label{eq_scaling}
\end{equation} 
with $r$ generically denoting all possible mass ratios $m/\msk$. The
indices $i,j=g,q,\bar{q}$ indicate the partonic initial state of the
 reaction $ij\to \stk\stkb$. The center-of-mass energy of the
partonic reaction, $\sqrt{s}$, is absorbed in the quantity $\eta =
s/4\msk^2-1$, which is better suited for analyzing the scaling
functions in the various regions of interest. We have identified the
renormalization and factorization scales: $\mu_R=\mu_F=\mu$. The
scaling functions are divided into the Born term $f^{B}$, the sum of
virtual and soft-gluon corrections $f^{V+S}$, the hard-gluon
corrections $f^{H}$, and the scale-dependent contributions $\bar{f}$.
The $\log^i \Delta$ ($i=1,2$) terms are separated from the soft-gluon
corrections and added to the hard-gluon part.  The hard-gluon
corrections are therefore independent of the cut-off for $\Delta \to
0$.\bigskip

\begin{figure}[ht]
\begin{center}
\epsfig{file=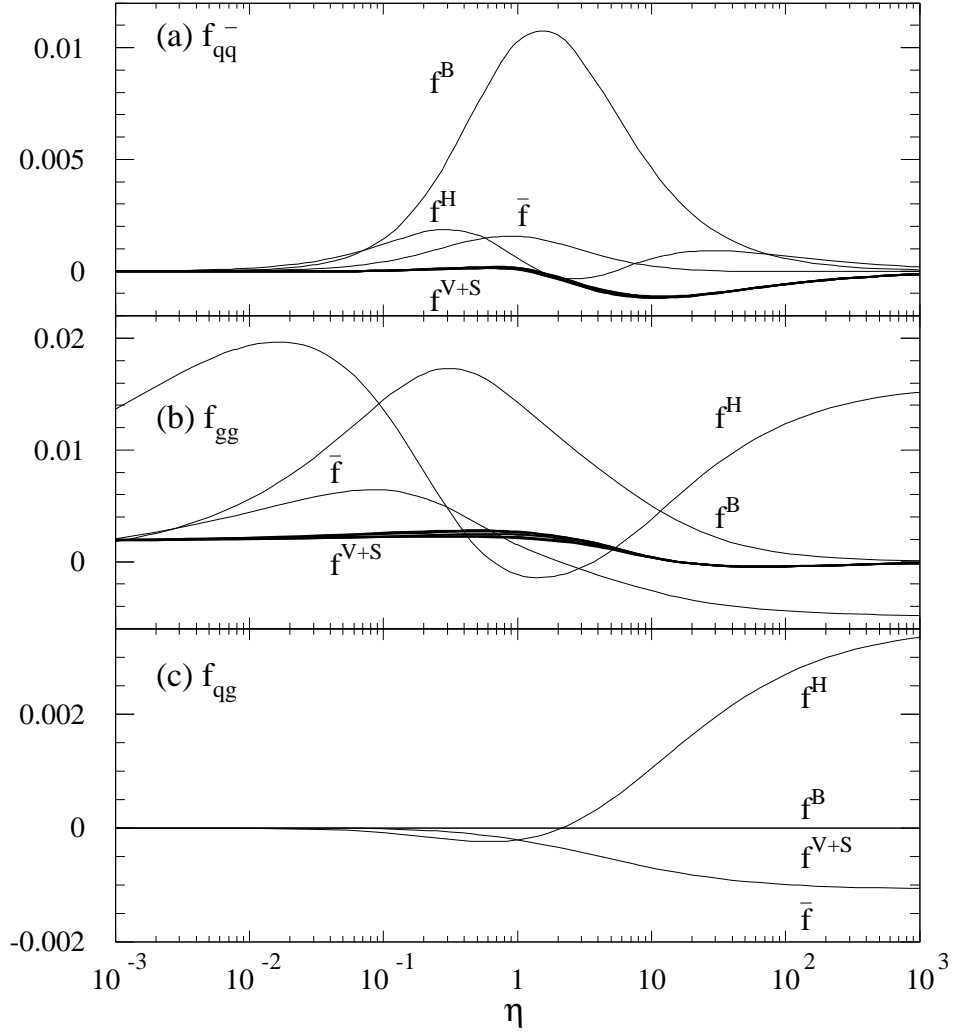,width=13cm}
\end{center}
\label{scaling}
\caption[]{\it The scaling functions for the production of 
           $\ste\steb$ pairs as a function of $\eta=s/4\mse^2-1$. 
           The notation follows Eq.\,(\ref{eq_scaling}). The 
           variation of 
           the scaling function $f_{ij}^{V+S}$ for all possible 
           values of the 
           mixing angle $\tmix$ is indicated by the line-thickness 
           of the
           corresponding curves. The scaling functions 
           $f_{g\bar{q}}$ are identical to $f_{qg}$.
           \label{fig_scaling}}
\end{figure}

The scaling functions are displayed in Fig.\,\ref{fig_scaling} for the
quark--antiquark, gluon--gluon, and (anti)quark--gluon channels. The
mixing angle and the gluino and light-flavor squark masses are defined
in a minimal supergravity scenario that will be discussed later in
detail. The dependence on these parameters is weak. In particular if
the mixing angle is varied over the full range, the impact on
$f^{V+S}$, the only scaling function affected by $\tmix$, is very
small. Due to the $\beta^3$ behavior of the LO quark--antiquark cross
section, not much structure is observed in this channel in the
vicinity of the production threshold ($\beta,\eta \ll 1$). However,
for the gluon--gluon channel two sources of large corrections can be
identified in the threshold region.  First, the exchange of
(long-range) Coulomb gluons between the slowly moving massive
particles in the final state leads to a singular Sommerfeld correction
$\sim \pi\as/\beta$, which compensates the LO phase-space suppression
$\beta$. The scaling function $f_{gg}^{V+S}$ therefore approaches a
non-zero value at the threshold. [For the $q\bar{q}$ incoming state
the cross section remains suppressed $\sim \beta^2$, since the LO
gluino-exchange contribution that dominates the threshold behavior of
light-flavor squark-pair production \cite{roland}, is absent for
stop-pair production]. It should be noted, however, that the screening
due to the non-zero lifetimes of the top-squarks reduces the Coulomb
effect considerably.  Second, as a result of the strong energy
dependence of the gluon--gluon cross sections near threshold, large
gluonic initial-state corrections of the type $\beta\log^i \beta$
($i=1,2$) emerge. The leading $\log^2 \beta$ terms are universal and
can be exponentiated. Near threshold, the scaling functions can be
expanded in $\beta$:
\begin{alignat}{2}
  &\\[-0.7cm]
  &f_{gg}^B =  \frac{7 \pi \beta}{384} &\qquad
  &f_{q\bar{q}}^B = \frac{\pi \beta^3}{54} 
\\
  &f_{gg}^{V+S}  =  f_{gg}^{B} \,\frac{11}{336 \beta} &\qquad
  &f_{q\bar{q}}^{V+S} =  -f_{q\bar{q}}^{B} \,\frac{1}{48 \beta} 
\nonumber\\
  &f_{gg}^{H} =  f_{gg}^{B} 
  \left[ 
   \frac{3}{2\pi^2}\log^2(8\beta^2)
  -\frac{183}{28\pi^2} \log(8\beta^2) 
  \right] &\qquad
  &f_{q\bar{q}}^{H}  =  f_{q\bar{q}}^{B} 
  \left[ 
   \frac{2}{3\pi^2}\log^2(8\beta^2) 
  - \frac{107}{36\pi^2}\log(8\beta^2) 
  \right] 
\nonumber\\
  &\bar{f}_{gg} = -f_{gg}^B \,\frac{3}{2\pi^2}\log(8\beta^2)  &\qquad
  &\bar{f}_{q\bar{q}} = -f_{q\bar{q}}^B \,\frac{2}{3\pi^2}\log(8\beta^2)
\nonumber
\end{alignat}
At high energies the NLO partonic cross sections in the gluon--gluon
and (anti)quark--gluon channels approach non-zero limits
asymptotically, to be contrasted with the scaling behavior $\sim1/s$
of the LO cross sections.  This is caused by the nearly on-shell
exchange of space-like gluons, associated with the inelastic Compton
processes and gluon radiation in the fusion process.  Exploiting the
factorization in transverse gluon momentum at high energies, the
high-energy scaling functions can be determined analytically:
\begin{alignat}{3}
  \\[-0.7cm]
  f_{gg}^H  &=&  \frac{2159}{43200\pi} &\qquad 
  f_{qg}^H  &=&  \frac{2159}{194400\pi} \\
 \bar{f}_{gg} &=&  -\frac{11}{720\pi} &\qquad
 \bar{f}_{qg} &=&  -\frac{11}{3240\pi} \nonumber
\end{alignat}
The scaling functions $f_{g\bar{q}}$ are identical to $f_{qg}$.  The
ratio of the $f_{gg}$ and $f_{qg}$ scaling functions is given by
$9:2$, corresponding to the probability for emitting a soft gluon from
a gluon (twice) or a quark.

The numerical analyses of the hadronic cross sections have been
performed for the Fermilab Tevatron $p\bar{p}$ collider with a
center-of-mass energy of $\sqrt{S}=1.8$~TeV, and for the CERN Large
Hadron Collider (LHC) with a $pp$ center-of-mass energy of
$\sqrt{S}=14$~TeV. We have adopted the CTEQ4M parametrization of the
parton densities \cite{cteq}.  The uncertainty due to different
parametrizations of the parton densities in NLO is less than $\sim
5\%$ at the Tevatron and less than $\sim 10\%$ at the LHC
(irrespective of the choice for the scale $\mu$). The difference
between the two estimates can be attributed to the fact that the
experimentally well-determined valence quarks dominate at the
Tevatron, while the gluons are more prominent at the LHC.  \bigskip

\begin{figure}[ht]
\begin{center}
\epsfig{file=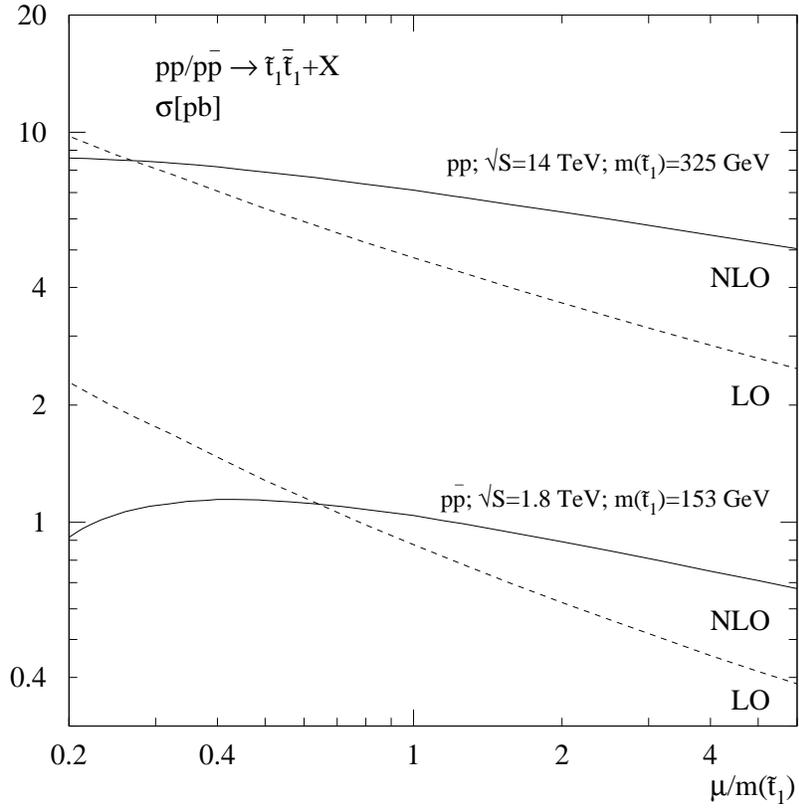,width=11cm}
\end{center}
\caption[]{\it Renormalization/factorization-scale dependence of the 
           total 
           cross sections for $\ste$-pair production at the Tevatron 
           and the 
           LHC. The SUSY mass parameters correspond to the central 
           values of 
           the SUGRA-inspired scenario described in the text. 
\label{fig_scale}}
\end{figure}

In Fig.\,\ref{fig_scale} we present the dependence of the total cross
sections for $\ste\steb$ production on the renormalization and
factorization scale $\mu=\mu_R=\mu_F$. For a consistent comparison of
the LO and NLO results, we have calculated all quantities
[$\as(\mu_R^2)$, the parton densities, and the partonic cross
sections] in LO and NLO, respectively.  In LO the scale dependence is
steep and monotonic: changing the scale from $\mu=2\mse$ to
$\mu=\mse/2$, the LO cross section increases by more than 100\%. In
NLO the scale dependence is strongly reduced, to about 30\% in this
interval at the Tevatron. At the same time the cross section is
considerably enhanced at the central scale ($\mu=\mse$). The results
are qualitatively similar for the LHC.

The magnitude of the SUSY-QCD corrections is illustrated by the $K$
factors in Table\,\ref{tab_kfac} for Tevatron and LHC energies. The
$K$ factor is defined as $K=\sigma_{NLO}/\sigma_{LO}$, with all
quantities calculated consistently in lowest and in next-to-leading
order. In these calculations the scale is fixed at the central value
($\mu=\msk$). For illustration, the supersymmetric parameters have
been fixed within the minimal supergravity (SUGRA) model\footnote{For
  an approximate solution of the renormalization-group equations in
  the SUGRA-inspired model, the program SPYTHIA \cite{spythia} has
  been adopted. For the Tevatron (LHC) parameters the input values
  $m_0=100$~GeV, $m_{1/2}=100\ (250)$~GeV, $A_0=300$~GeV,
  $\tan\beta=1.75$, and $\mu>0$ have been chosen, generating the mass
  and mixing parameters quoted above.} such that $\mg=284\ (627)$~GeV
and $\sz =-0.99\ (-0.94)$ for the Tevatron and the LHC, respectively.
The stop masses in these scenarios are given by $\mse=153\ (325)$~GeV
and $\msz=347\ (592)$~GeV, all other squarks are assumed to be mass
degenerate with $\ms=256\ (539)$~GeV.  However, in order to focus on
the mass dependence of the cross sections and of the NLO corrections,
the mass of the produced stop particles is varied around the
SUGRA-inspired central value, independently of the other SUSY
parameters. In the mass range considered, the SUSY-QCD corrections are
small (and negative) if the $q\bar{q}$ initial state dominates,
cf.~Table~\ref{tab_kfac}. If, in contrast, the $gg$ initial state
dominates, the corrections are positive and reach a level of 30--40\%.
The relatively large mass dependence of the $K$ factors for
$\ste\steb$ production at the Tevatron can therefore be attributed to
the fact that the $gg$ initial state is important for small $\mse$,
whereas the $q\bar{q}$ initial state dominates for large $\mse$.
\begin{table}[ht]
\begin{center}
\begin{tabular}{|cc||c|c||c|c|}
\hline \rule[0mm]{0mm}{5mm}
    & $\mst$ [GeV]
    & $K_{\rm Tevatron}$
    & $gg_{\rm in}$ : $q\bar{q}_{\rm in}$
    & $K_{\rm LHC}$
    & $gg_{\rm in}$ : $q\bar{q}_{\rm in}$ \\[2mm] \hline
      $\ste \steb$ \rule[0mm]{0mm}{5mm}
    & 70$\hphantom{0}$
    & 1.41
    & 0.64 : 0.36
    & 1.25
    & 0.96 : 0.04 \\
    & 110
    & 1.30
    & 0.41 : 0.59
    & 1.32
    & 0.95 : 0.05 \\
    & 150
    & 1.19
    & 0.25 : 0.75
    & 1.37
    & 0.94 : 0.06 \\
    & 190
    & 1.11
    & 0.15 : 0.85
    & 1.40
    & 0.92 : 0.08 \\[1mm] \hline
      $\stz\stzb$ \rule[0mm]{0mm}{5mm}
    & 280
    & 1.01
    & 0.06 : 0.94
    & 1.44
    & 0.89 : 0.11 \\
    & 320
    & 1.00
    & 0.04 : 0.96
    & 1.45
    & 0.88 : 0.12 \\
    & 360
    & 0.98
    & 0.03 : 0.97
    & 1.46
    & 0.86 : 0.14  \\
    & 400
    & 0.95
    & 0.02 : 0.98
    & 1.48
    & 0.85 : 0.15 \\[1mm] \hline
\end{tabular}
\end{center}
\caption[]{\it $K$ factors for diagonal stop-pair production at 
 the Tevatron 
 and the LHC for a sample of stop masses. Scale choice: $\mu=\mst$. 
 The SUGRA-inspired parameters adopted in the 
 calculation of the higher-order corrections are defined in the 
 text. For completeness also the LO initial-state $gg$ and $q\bar{q}$
 fractions are given.\label{tab_kfac}}  
\end{table}

\begin{figure}[ht]
\begin{center}
\epsfig{file=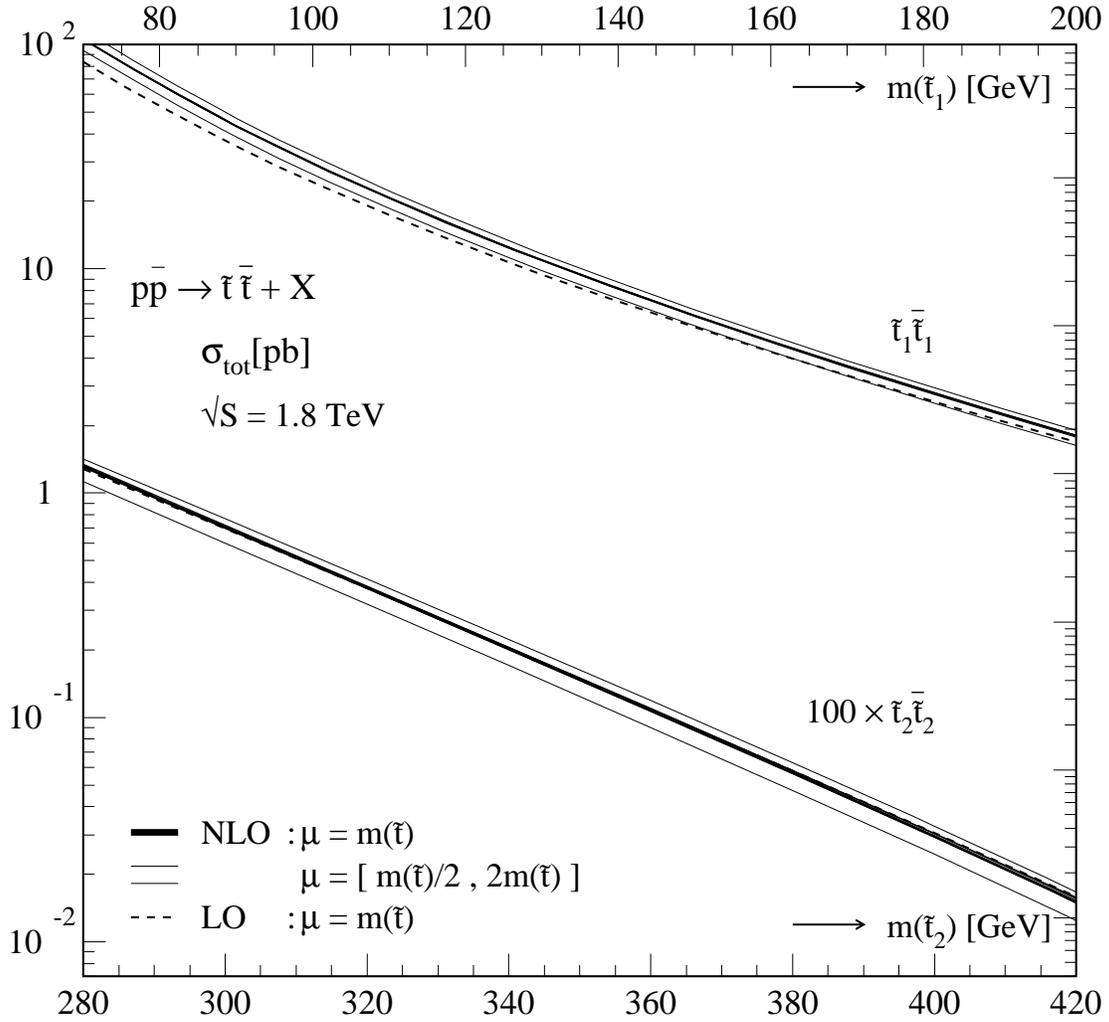,width=15cm}
\end{center}
\caption[]{\it The total cross sections for the production of pairs 
           of stop
           particles ($\,\stk \stkb$) at the Tevatron as a function 
           of the 
           stop masses. The band for the NLO result indicates 
           the 
           uncertainty due to the renormalization/factorization 
           scale. The 
           light-flavor squark masses, the gluino mass and the 
           mixing 
           parameter are derived within the SUGRA-inspired scenario 
           defined in 
           the text. The line-thickness of the NLO curves represents 
           the 
           simultaneous variation of the gluino mass between 
           200\ (284) and 
           800~GeV for $\ste(\stz)$-pair production and the 
           variation of 
           $\sz$ over its full range.\label{fig_sig1}}
\end{figure}

\begin{figure}[ht]
\begin{center}
\epsfig{file=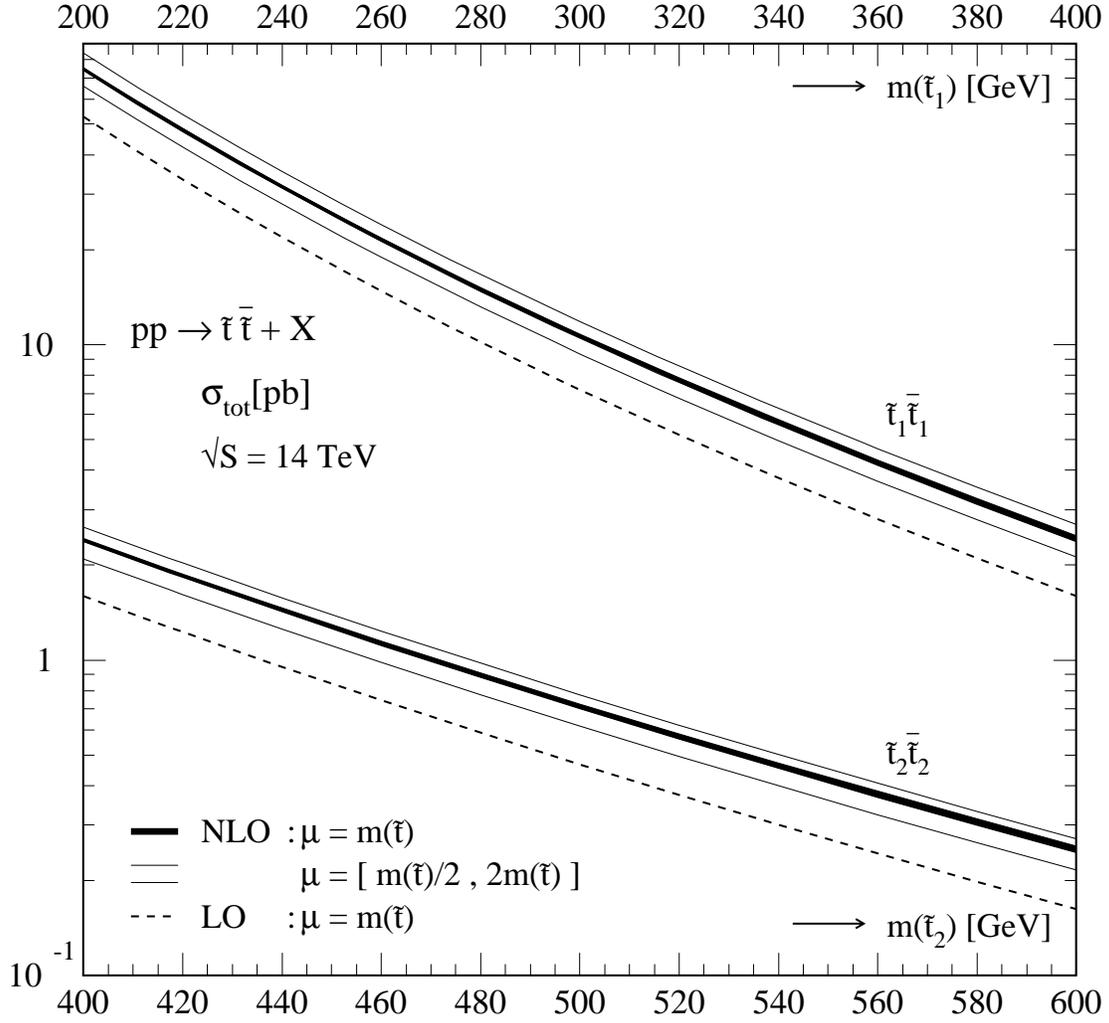,width=15cm}
\end{center}
\caption[]{\it The same as Fig.\,\protect\ref{fig_sig1}, but 
           for the LHC.
           The SUSY mass spectrum is described in the text. The gluino 
           mass is varied between 400\ (600) and 900~GeV for 
           $\ste(\stz)$-pair
           production.\label{fig_sig2}}
\end{figure}

The total cross sections play a crucial role in the experimental
analyses. They either serve to extract the exclusion limits for the
mass parameters from the data, or, in the case of discovery, they can
be exploited to determine the masses of the stop particles. The total
cross sections for $p\bar{p}/pp \to \ste\steb, \stz\stzb$ are given in
Figs.\,\ref{fig_sig1}/\ref{fig_sig2} for the Tevatron/LHC,
respectively. The cross sections depend essentially only on the masses
of the produced stop particles, and very little on the other
supersymmetric parameters, {\it i.e.}~the gluino mass, the masses of
the light-flavor squarks and the mixing angle. The variation of the
cross section with the gluino mass and the mixing angle is indicated
by the thick NLO curves.  With cross sections typically in the range
between 1 and 100 pb, sufficiently large samples of $10^3$ to $10^5$
stop events can be accumulated at the Tevatron for an integrated
luminosity of $\int {\cal L}=1\;\rm{fb}^{-1}$, provided these
particles exist in the mass range below 200 GeV. For an integrated
luminosity of $\int {\cal L}=100\; \rm{fb}^{-1}$ at the LHC a large
sample of $10^5$ to $10^7$ stop events could be collected for masses
in the 200--500~GeV range. \bigskip

\section[]{Mixed $\ste\stzb$ and $\stz\steb$ pairs}
 
In $p\bar{p}/pp$ collisions, the mixed final states $\ste\stzb$ and
$\steb\stz$ cannot be produced in lowest order.  The production cross
section for non-diagonal stop pairs is therefore of order
$\alpha_s^4$. This higher-order cross section is small but complicated
to calculate. We therefore discuss the size of the cross section in
the limit of large gluino mass. In this limit, the four-squark
couplings relevant for non-diagonal stop-pair production are described
by the Lagrangean
\begin{alignat}{7}
  &{\cal L}'_4 \quad &=\ & \frac{g_s^2}{4}\,\sz\,
      \big( \ste{}_i^*\,\stz{}_j + \stz{}_i^*\,\ste{}_j \big)\,
      \bigg\{ \cz\,A_2^{ijkl}\,\big( \ste{}_k^*\,\ste{}_l 
                                   - \stz{}_k^*\,\stz{}_l \big) 
       \notag \\
  &&& \hphantom{\frac{g_s^2}{4}\,\sz\,\big( \ste{}_i^*\,\stz{}_j 
                + \stz{}_i^*\,\ste{}_j \big)\, A}
          {}+ 2\,A_1^{ijkl} \sum_{\sq\neq\st} 
              \big( \sql{}_k^*\,\sql{}_l - \sqr{}_k^*\,\sqr{}_l \big)
      \bigg\}   
\end{alignat}
[where however the second term does not contribute to the cross
section for mass degenerate left/right light-flavor squarks $\sq$].
Only two one-loop diagrams contribute to the scattering amplitude in
this limit (see Fig.\,\ref{fig_feyn}c).  They involve the production
of diagonal $\stk\stkb$ pairs in $gg$ collisions, subsequently
transformed to non-diagonal $\ste\stzb$ and $\stz\steb$ pairs by
rescattering in the final state. The loops can easily be evaluated:
\begin{eqnarray}
\hat{\sigma}_\infty [gg\to\ste\stzb+\steb\stz] = 
\sin^2(4\tmix)\,\frac{37}{13824}\,
\frac{\alpha_s^4\,\lambda^{1/2}}{2 \pi s^3}\, 
\left| \mse^2 \log^2(x_1) - \msz^2 \log^2(x_2)
\right|^2 &&\quad
\end{eqnarray}
where the subscript in the cross section $\hat{\sigma}_\infty$
indicates the limit $\mg \to \infty$. The coefficient $\lambda^{1/2}$
is the usual 2-particle phase-space factor, {\it i.e.}
$\lambda=[s-(\mse+\msz)^2][s-(\mse-\msz)^2]/s^2$, and $x_k =
(\beta_k-1)/(\beta_k+1)$; the logarithmic singularities are defined
properly by the infinitesimal shift $s \to s + i\varepsilon$ in
$\beta_k$.

The cross section depends strongly on the mixing angle $\tmix$ through
the overall factor $\sin^2(4 \tmix)$. Numerical values for the
diagonal and non-diagonal pair cross sections are compared in
Table\,\ref{tab_sig} for the default SUGRA-inspired SUSY parameter
sets \cite{spythia} adopted already earlier. Note that the mixed-pair
cross section is given in this table {\it without the mixing factor\/}
$\sin^2(4 \tmix)$. Evidently, the values for the cross section for
producing mixed stop pairs in the large $\mg$ limit are very small at
the Tevatron as well as at the LHC.
\begin{table}[ht]
\begin{center}
\begin{tabular}{|c|l||cc||cc|}
\hline \rule[0mm]{0mm}{5mm}
    & $\sigma$[fb]
    & $\sigma_{q\bar{q}}$
    & $\sigma_{q\bar{q}}^{\rm limit}$
    & $\sigma_{gg}$
    & $\sigma_{gg}^{\rm limit}$ \\[1mm] \hline
      Tevatron
    & $\ste \steb$ \rule[0mm]{0mm}{5mm}
    & 0.64 $\cdot 10^{3}$
    & 0.64 $\cdot 10^{3}$
    & 0.42 $\cdot 10^{3}$
    & 0.42 $\cdot 10^{3}$    \\
    & $\stz \stzb$ \rule[0mm]{0mm}{5mm}
    & 1.51 
    & 1.54 
    & 0.105 
    & 0.108  \\
    & $\ste \stzb + \stz \steb$ \rule[0mm]{0mm}{5mm}
    & --
    & 0
    & --
    & 1.54 $\cdot 10^{-4}$ \\[1mm] \hline
      LHC
    & $\ste \steb$ \rule[0mm]{0mm}{5mm}
    & 0.60 $\cdot 10^{3}$
    & 0.59 $\cdot 10^{3}$
    & 6.50 $\cdot 10^{3}$
    & 6.47 $\cdot 10^{3}$    \\
    & $\stz \stzb$ \rule[0mm]{0mm}{5mm}
    & 37.1
    & 37.1
    & 0.23 $\cdot 10^{3}$
    & 0.24 $\cdot 10^{3}$    \\
    & $\ste \stzb + \stz \steb$ \rule[0mm]{0mm}{5mm}
    & --
    & 0
    & --
    & 3.53 $\cdot 10^{-2}$   \\[1mm] \hline
\end{tabular}
\end{center}
\caption[]{\it Cross sections for diagonal and non-diagonal pair 
           production 
           at the Tevatron and the LHC, using the default 
           SUGRA-inspired
           values for the SUSY parameters. The non-diagonal 
           results are given 
           without the mixing factor $\sin^2(4 \tmix)$. Scale 
           choice: average
           mass of the produced stop particles. The superscript 
           'limit'
           denotes the asymptotic value of the cross section 
           for large gluino masses. \label{tab_sig}}   
\end{table}
\bigskip

\section[]{Summary} 

The picture that has emerged from the SUSY-QCD analysis, is quite
simple. (i) The cross sections for the production of diagonal pairs
$p\bar{p}/pp \to \ste\steb,\stz\stzb$ depend essentially only on the
masses of the stop particles produced and very little on the other
supersymmetric parameters.  Bounds on the $\ste\steb$ production cross
section can therefore easily be translated into lower bounds on the
lightest stop mass without reference to other supersymmetric
parameters. On the other hand, if stop particles were to be
discovered, the cross section can be exploited directly to determine
the two stop masses $\mse$ and $\msz$. (ii) If mixed stop pairs could
be discovered in $pp \to \ste\stzb+\stz\steb$ for sufficiently high
integrated luminosity, the mixing angle can be derived from the
magnitude of the cross section which is proportional to $\sin^2(4
\tmix)$. In contrast to mixed stop-pair production via $Z$ exchange in
$e^+e^-$ annihilation \cite{lincol}, the cross section is suppressed
by ${\cal O}(\alpha_s^2)$ with respect to diagonal pair production.
\bigskip

\noindent {\bf Acknowledgements:}
We are very grateful to S.~Lammel for advise and discussions on the
search for stop particles at the Tevatron. Collaboration with
R.~H\"opker is also gratefully acknowledged. W.B., M.K. and M.S. thank
the DESY Theory Group for the warm hospitality extended to them during
a visit.\bigskip

\noindent {\bf Note:} The program for calculating the 
stop cross sections has been included into the package {\tt PROSPINO},
http://wwwcn.cern.ch/$\sim$mspira.

\newpage

\end{document}